# Tertiary EOR-like microfluidic experiments: influence of viscosity ratio on oil clusters mobilization

**Haohong Pi [a], Abdelaziz Omari [a*], Giuseppe Sciumè [a,b]**

a. University of Bordeaux, CNRS, Bordeaux INP, I2M, UMR 5295, F-33400, Talence, France

b. Institut Universitaire de France (IUF)

**\*** Corresponding author address email: omari@enscbp.fr

## Abstract

Understanding the pore-scale dynamics of immiscible two-phase flow in porous media is crucial for optimizing EOR strategies. In this work, we investigate the mobilization dynamics of oil clusters by means of microfluidic devices that allow pore scale direct characterization of flow in water-wet chips. We varied both flow rates during waterflooding and the viscosity ratio by injecting Glycerol/water mixtures of various compositions right after the waterflooding period. During waterflooding, the flow rate has only a limited impact on residual oil. With a subsequent injection of a Glycerol/water mixture, the oil recovery is significantly enhanced. To better understand the recovery mechanisms, oil clusters were categorized into droplets, blobs and ganglia. Increasing the viscosity of the injected mixture resulted in only a slight reduction in the number of ganglia but significantly decreased their total volume, thus reducing overall oil saturation. This is due to ganglia breakup into smaller ganglia, blobs and droplets that are subsequently mobilized and transported away, while remaining parts of original ganglia still remain trapped. As long as droplets and blobs are considered, their number is seen to only weakly change by the increase of mixture viscosity and even their number may temporarily increase as they result from ganglia rupture. So, the process can be separated in two main steps: ganglia breakage that feed the medium in blobs and droplets and a second step where such moving oil entities are transported. The characteristic time for oil transport is believed to be longer than that required for ganglia breakage.

# Introduction

A mechanistic understanding of pore-scale dynamics of immiscible two-phase flow through porous media is essential for various subsurface operations, including enhanced oil recovery (EOR) [1–4], $CO_2$ geo-sequestration [5–7] and soil pollution remediation [8,9]. Although these mechanisms were initially developed for EOR due to historical priorities, the knowledges acquired in EOR are now currently being applied to the two other application fields [10,11]. So, in the context of EOR, improving oil recovery rates requires a deep understanding of oil displacement mechanisms, particularly in secondary and tertiary recovery processes. To that end, fluids with varying compositions and rheological properties are typically injected into oily reservoirs to modify the interfacial tension, viscosity of fluids or wettability of the reservoir [12–20].

On the laboratory scale, the assessment of EOR methods traditionally relies on conventional core-flood experiments, which provide quantitative and reliable oil recovery data at the Darcy scale (e.g., relative permeabilities of fluids and recovery rate of the original oil in place, OOIP, among others) [21–23]. However, a significant limitation of core-flood experiments is inherent to their "black box" nature, as they do not allow direct observation of critical pore-scale phenomena (e.g., snap-off, pore filling, etc.) or the mobilization of remaining oil. This is why microfluidic micromodel devices are now extensively used. These on-chip experimental devices allow hence direct observation of two-phase flow at pore scale by dyeing fluids and using microscopy techniques, as optical and fluorescence microscopy imaging, and are sometimes coupled with particle image velocimetry techniques [3,24–33]. The microchips used are fundamentally transparent, 2D synthetic porous media designed to replicate the structure of the actual porous media of interest. Furthermore, to overcome their lack of representativeness of true 3D porous media, 2.5D chips have recently become available [29]. Anyway, despite this limitation, their most significant advantage lies in their ability to enable real-time visualization of fluid flow, providing insights into the fundamental mechanisms driving the movement of wetting and non-wetting phases at the pore scale.

The fundamentals of two-phase flow in porous media build upon the pioneering work of Saffman and Taylor [34], who theoretically and experimentally studied the instability of the interface between two immiscible fluids in a Hele-Shaw cell. They demonstrated that the becoming of such instability is primarily influenced by the interfacial tension of the two fluids and their viscosity and density. For incompressible fluids, the flow structure depends on the capillary number (Ca) that is the ratio of viscous forces to capillary forces, as well as viscosity ratio (M) between invading and defending phases. Lenormand et al. [35] later have extended this work to explore the dynamics of two immiscible fluids in porous media, summarizing their findings in a (Ca, M) graphical phase diagram with three fundamental flow regimes: At high Ca and favourable M (M>1) a piston-like flow with a stable displacement that arise in minimum residual saturation. In contrast, viscous fingering occurs when M<1, with multiple thin fingers extending toward the outlet, causing rapid breakthrough and high residual saturation. At very low Ca, however, capillary fingering produces fingers that grow in all directions, resulting in intermediate residual saturation.

On the experimental side, numerous studies [2,3,29,31,32,36–42] have been conducted on both 2D and 3D porous media. Some of these studies [31,37,40,41] used 2D microfluidic chips with constant depth to quantitatively investigate changes in fluid saturation at various (Ca, M) couples through microscopy and image analysis. Their work provided detailed insights into flow structures, revealing transitions between the three fundamental regimes, thereby refining the Lenormand's findings. Furthermore, Hu et al. [38] have investigated two-phase fluid displacement in a 3D glass bead column and reported results closely resembling those obtained in on-chip experiments, suggesting that the behaviour observed in 2D micromodels can be representative of what might happen in 3D porous media, but only when capillary number is defined with care [42,43]. Specially, since permeability in 2D micromodels is primarily determined by the etching depth of the channels, while in 3D core systems it is more closely related to pore network geometry, applying capillary number definitions from 3D models to 2D systems can lead to inconsistencies. These discrepancies were extensively discussed by Tang et al. [43], who proposed a modified Ca based on 2D micromodels to align results from 2D more closely with those from 3D experiments.

Moreover, in-depth research at the pore scale has provided critical insights. Key displacement events, such as Haines jumps, snap-off, and pore-filling [36,39,46-48], were extensively analysed to understand their roles in displacement dynamics. Crucially, the mobilization of trapped oil clusters has proven vital for enhancing oil recovery in EOR operations, prompting a range of experiments specifically designed to explore the transport and behaviour of these clusters within porous media [3,4,13,18,23,45–47]. Oil clusters can deform, move, or break into smaller clusters (e.g., droplets, blobs) based on fluid properties and flow strength [1,48–50]. Some studies [51–54] indicate that large ganglia break up as Ca increases, with a fragmentation rate that rises alongside Ca, while the resulting clusters, whose sizes influenced by upstream pore structures, become mobilized and removed only at high Ca [27,44]. Alternatively, other research suggests that ganglia do not break up and only mobilize and move when they exceed a threshold size [1,55–57], with their mobilization potential influenced by ganglia size [46,58–61]. Some studies report that larger or elongated ganglia are more easily mobilized due to greater local pressure differences [55,59,62], while others suggest that elongated ganglia resist mobilization [63], with small ganglia generally showing higher transport rates [1,64–66]. Furthermore, Zarikos et al. [24] used microscopic particle tracking velocimetry to show quantitatively that droplets remain unaffected by pressure differences, while blobs mobilize when momentum transfer is sufficient, and ganglia rupture in non-flow area at high Ca, mobilizing parts of the ganglion. Notably, current research have reported that the displacement characteristics of clusters in 2D and 3D porous media are comparable [67].

Nevertheless, nearly all of these typical EOR studies have principally concerned the secondary period and limited investigations were devoted to the tertiary period, during which a viscous fluid is usually injected after waterflooding. This work primarily aims to address this gap by investigating the mobilization dynamics of oil clusters in the tertiary period using a well-controlled on-chip microfluidic experimental setup. To achieve this, we systematically varied the viscosity of the injected fluid after sequential water saturation, oil drainage, and waterflooding. Our main goal was to characterize in detail the process by which more oil is produced by this way. We focus here on investigating the dynamics of removal of retained oil clusters depending on injected fluid viscosity and flow rate. Particular attention will be paid to determining which type of oil clusters are mobilized by following their size and location over time. In the next section, the experimental setup and procedure are presented. This section also provides details on how raw data are analysed. In the subsequent section, experimental results are presented and discussed within the framework of knowledge from the literature. In particular, we successively present both drainage and waterflooding steps before describing more extensively the consequences of the subsequent injection of Glycerol/water mixtures of various viscosities. In that respects, we propose a phenomenological description of oil clusters breakage and mobilisation processes that lead to enhancing the oil recovery. The paper end with drawing some conclusions and perspectives.

## 2 Materials and Methods

### 2.1 Fluids

The oil and water used in the experiment were commercial rapeseed oil and deionized water containing 10 g/L NaCl, respectively, with both fluids filtered prior to use. Aqueous mixtures of salted water and glycerol (≥99.0% GC, Sigma-Aldrich) were prepared by gently mixing pre-defined amounts of salted water with glycerol to achieve glycerol contents ranging from 0 to 82% (w/w). The rapeseed oil was dyed with Oil Red O (Sigma-Aldrich) at a concentration of 0.01 wt%, while the salted water and Glycerol/water mixtures were respectively dyed with Ecoline Vert Foncé 602 (0.2 wt%) and Ecoline Bluish Violet 548 (0.2 wt%), (both are Royal Talens, Netherlands). The density ($\rho$) and viscosity ($\mu$) of the all fluids were measured using gravimetric methods and an Ostwald viscometer, respectively. The interfacial tension (IFT, $\sigma$) between the oil and water and between the oil and the Glycerol/water mixtures was assessed using a Force Tensiometer (KRÜSS, Germany). The results revealed that the IFT between the oil and the water or Glycerol/water mixtures remained almost constant, at approximately 55 mN/m. All the measurements were performed at a room temperature (20 ± 1 °C), and the obtained fluid characteristics are summarized in **Table 1**.

*Table 1. The properties of experimental fluids*

|  | Oil | Water | Mixtures of Glycerol/water (w/w) | | | |
| --- | --- | --- | --- | --- | --- | --- |
|  |  |  | 46/54 | 54/46 | 68/32 | 82/18 |
| **Density (kg/m$^3$)** | 910 | 1000 | 1109 | 1137 | 1175 | 1214 |
| **Viscosity (mPa·s)** | 80 | 1 | 4.48 | 7.82 | 19.4 | 79.95 |

Additional chemicals, including absolute ethanol, sodium hydroxide (NaOH, 20% w/v solution, Scharlau, Spain), and hydrochloric acid (HCl, 37%, Scharlau, Spain), as well as the surfactant Enordet O332 (%AM = 29.55, Shell Chemicals, Amsterdam), were used to clean the microfluidic chip.

### 2.2 Microfluidic Chip

#### 2.2.1 Geometry and Physical Properties

The microfluidic chip was water-wet with a rock-like network supported by Micronit Microtechnologies (Netherlands). The chip, supplied in a black polypropylene box, is fabricated from borosilicate glass using isotropic etching to create the porous matrix and with the inlet and outlet channels that enable flow distribution, as shown in **Fig. 1(a)**. The porous matrix consists of a network of throats and pores formed by irregularly shaped solid grain pillars. The etching depth is 20 μm, while the overall plane size of the porous area is 20 mm in length, 10 mm in width, which constitutes the area of interest in the experiment. The pore volume (PV) was 2.4 μL and the porosity ($\phi$) was of approximately 0.6. The measured 1D absolute permeability, $k$, of the microfluidic chip was 2.7 Darcy ($\approx 2.66 \cdot 10^{-12}$ m²). Moreover, the original pore size distribution, shown in **Fig. 1(b)**, has been obtained in this study using an image-treatment algorithm known as Sub-Network of the Over-Segmented Watershed (SNOW) method, which generates a pore network model based on the segmented pore space, representing pores as circular nodes [68–70].

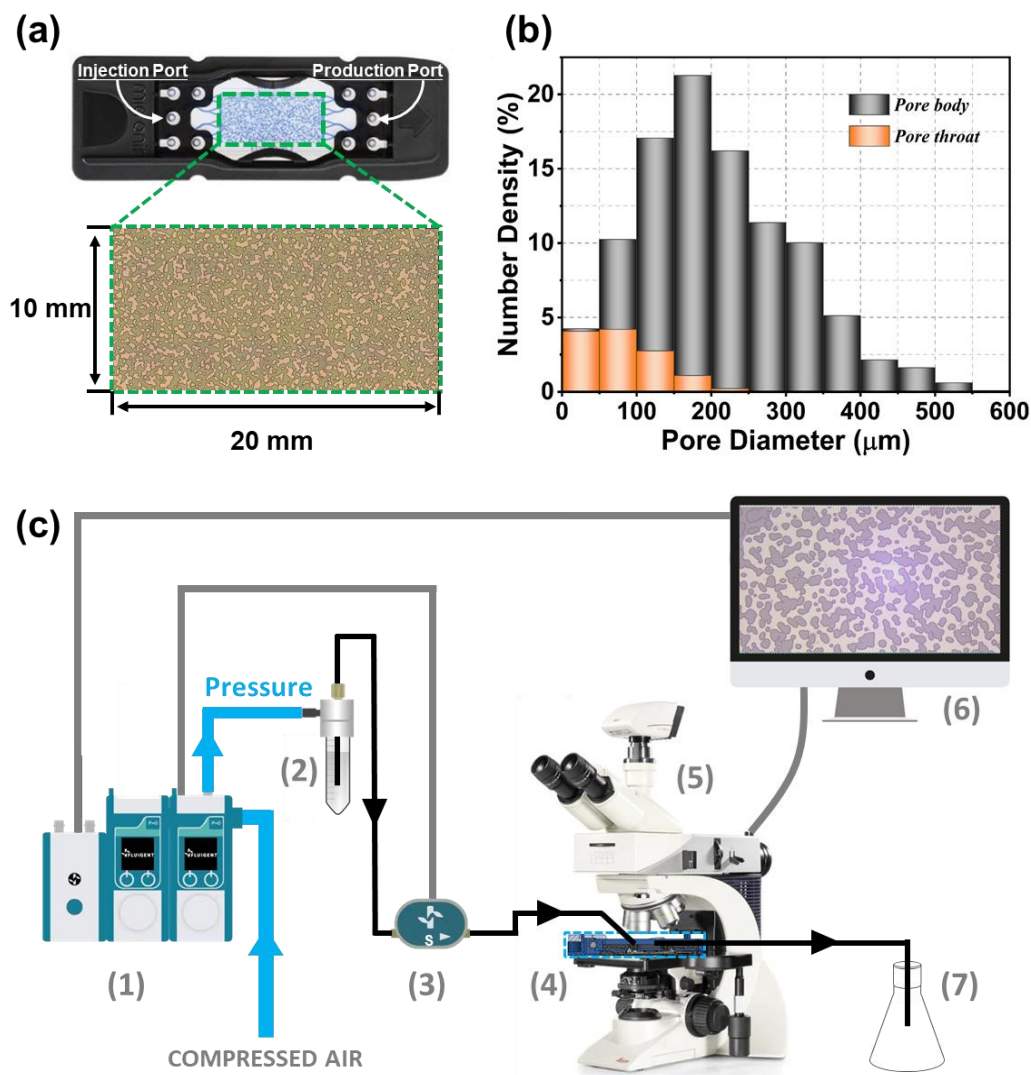

***Fig. 1.*** *(a) The microfluidic chip and the dimensions and pore network structure of the porous matrix. (b) The original pores size distribution of pore body and throat. (c) Schematic illustration of the microfluidic setup, which consists of the following components: (1) pressure controller pump; (2) P-CAP reservoir; (3) flow sensor (Flow Unit); (4) microfluidic chip holder with a fixed chip; (5) microscope with digital camera; (6) lab computer software system; (7) container for fluid recovery. Blue lines represent pneumatic tubes, black lines indicate microfluidic tubes (diameter 0.22 mm), and grey lines correspond to cables.*

2.2.2 Microfluidic chip Cleaning Procedure

Before each experiment, the microfluidic chip was thoroughly rinsed with 10 PV of pure water, followed by drying it with compressed nitrogen. After each experiment, an extensive cleaning protocol was performed to fully restore the microfluidic chip to its initial state. This process does respectively involve rinsing with: surfactant, deionized water, absolute ethanol, deionized water, NaOH solution, HCl solution, and finally deionized water to ensure a complete clean [71]. The surfactant (5 wt%) and absolute ethanol were used to remove any remaining non-wetting and wetting fluids. Other stubborn residues were dissolved by a 5 w/v% NaOH solution, and then neutralized by a 5% HCl solution. More than 10 PV fluids were injected at 1000 mbar during each rinse. After the last rinse with deionized water to ensure thorough cleaning, the micromodel was dried by injecting pressurized dried air at 2000 mbar via a pressure generator. Finally, the chip surfaces were cleaned with absolute ethanol, then dried again with pressurized air before being sealed for storage.

*2.3 Experimental Set-up*

The illustration of the overall microfluidic setup is shown in **Fig. 1(c)**. The setup consists of a pressure control system, the imaging set-up, the previously described microfluidic chip (which is mounted in a chip holder) a container for the recovery of fluids and the needed connectivity (cables and tubes).

The pressure control system, provided by FLUIGENT, consists of a pressure controller pump (1) which takes the pressurized air from a compressor and inject it at a given pressure (within a range of 0 to 2000 mbar) within a reservoir (2) containing the fluid. The pressurized air pushes out the fluid within a microfluidic tube (in black in **Fig. 1(c)**). The fluid within the microfluidic tube passes through a flow sensor (3) that measures flow and thanks to the interconnection with the pressure pump (1) enables to work at a prescribed flow rate. When we work at a prescribed flow rate (within a range from 0 to 8 μL/min), given the continuous information provided by the flow sensor (3), the pressure pump (2) continuously adjusts the pressure to maintain the prescribed flow rate. The pressure pump is connected to a computer software (6) that records the data and control the equipment. This system enables precise, low flow rate flooding of the microfluidic chips during experiments.

The microfluidic chip is secured in a holder (4) that connects it to the injection and production tubing for fluid injection and extraction. A Leica DM2700 M microscope (5) with magnifications ranging from 5× to 100× (22 mm field of view) is used to observe the flow of the fluids within the chip. Fluids from the P-CAP reservoir are injected into the chip, and the digital cameras mounted on the microscope, controlled by the computer (6), records the flow behaviour. An external collection container (7) is used to collect the fluids generated during the experiment.

*2.4 Experimental Procedures and Image Processing*

*2.4.1 Experimental Procedures*

Complete experimental sequences are schematically illustrated in **Fig. 2**. In general, the sequence consists of four successive steps: water saturation, oil drainage, waterflooding, and Glycerol/water mixture flooding. The sketches on the left side of the figure provide an artistic view of the system state at each step, while the right side presents the major experimental outputs, including fluid breakthrough, fluid saturations, and their relative parameters. The experiment proceeded following the steps:

*Water Saturation* – Before oil drainage, the microfluidic chip was first carefully saturated with water, with its porosity determined to be approximately 0.6 and its pore volume measured at ~ 2.4 μL. The absolute permeability ***k*** was then measured by flooding the chip at various flow rates and recording the corresponding pressure drops (ΔP) at steady state. The use of the Darcy law then gives a ***k*** value of 2.7 Darcy.

*Oil Drainage* – Subsequently, the dyed oil was injected at 0.1 μL/min for 6 PV to reach the irreducible water saturation ($S_{wi}$). At the end, the flow rate was then increased to 0.5 μL/min and 1.0 μL/min and injecting 6 PV of oil for each flow rate.

*Waterflooding* – The drainage here above was followed by a waterflooding step, during which dyed water was injected at Q = 0.1 μL/min for 6 PV to displace oil until reaching the residual oil saturation ($S_{or}$), at which point only water was being produced [72]. The flow rate was then varied as before in order to examine its impact on change in $S_{or}$. At that time, to check experimental reproducibility, these two previous steps were repeated 5 times separately. The results from these steps will be given later as averages with their corresponding standard deviations.

*Glycerol/water mixture flooding* – Following the previous waterflooding step, dyed Glycerol/water mixtures were injected at a Q of 1.0 μL/min to investigate the impact of the viscosity ratio $\eta_R$ on sweep efficiency enhancement and flow structure. Each particular mixture, was injected for a total volume of 18 PV. In a first case (control case), water injection was continued for 18 PV (extended waterflooding), while in the other four cases we injected Glycerol/water mixtures which compositions were listed in **Table 1**. During this tertiary recovery step, the definitions of the invading and defending fluids are a bit more delicate since we have three fluids in presence. Thus, we define another, more meaningful viscosity ratio ($\eta_R = \frac{\mu_{mixture}}{\mu_{water}}$) to represent the viscosifying rate of injected aqueous fluids. All experiments were performed at room temperature (20 ± 1°C) and each of them was visually monitored to capture images at key moments and right location for both qualitative and quantitative analysis, while ΔP and flow rate were continuously recorded.

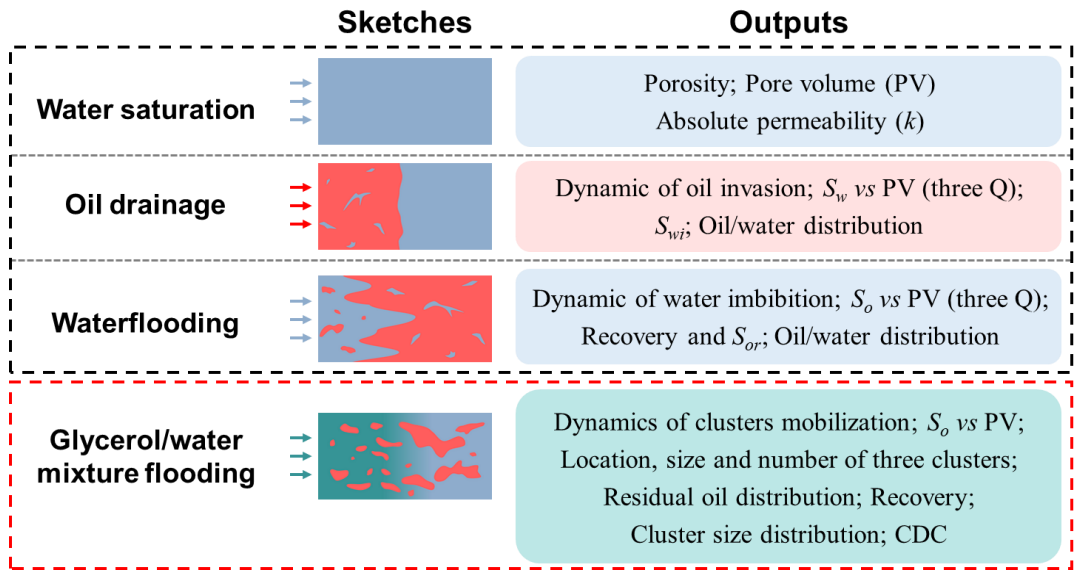

*Fig. 2. Experimental Procedures: Four sequential steps and the corresponding sketches that illustrate the typical fluid distribution during each step (excluding the solid phase): blue represents water, red represents oil, and green represents the Glycerol/water mixture. The measured and obtained results at each step are listed in outputs.*

*2.4.2 Image Acquired and Imaging Processing*

Images and video were acquired and recorded with a HD digital camera connected to the computer and controlled by LAS X software. Due to field-of-view limitations, each full-chip image was composed of 12 overlapping images (3840 × 2160 pixels each), as shown in **Fig. 3(a)** (the figures refer to waterflooding). These images were stitched together using ImageJ and then aligned to focus on the key area of interest, as shown in **Fig. 3(b)**. Although these images provide qualitative insights into fluids flow, they lack the precision required for quantitative analysis. To address this, further processing was performed to clearly differentiate and enhance the contrast between the wetting, non-wetting, and solid phases with color-coding: oil in red, water in blue, and solid grains in black, as shown in **Fig. 3(c)**. In case of mixtures flooding images (not shown here), mixtures were represented in dark blue. Subsequently, thresholding techniques were applied to identify each phase, allowing for precise measurements of oil cluster size, number, and distribution within the porous medium.

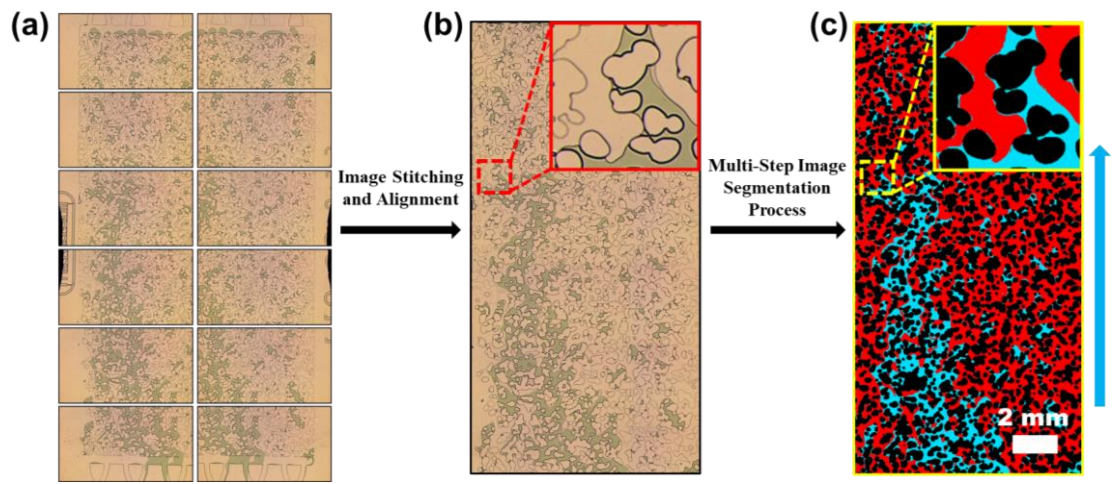

*Fig. 3. Image Acquisition and Processing: (a) The total 12 images of a full-chip image, individual images were taken with over 30% overlap between adjacent frames; (b) The resulting full-chip image; (c) The final processed image, presented in colour. The selected region highlighted to reveal features before (b) and after(c) processing. Blue arrows indicate the fluid flow direction.*

Moreover, to accurately quantify and analyse flow behaviour, as well as the change of the location and size of each phase, it was necessary to capture as many images as possible during the experiment. In this work, the acquisition frequency ($\Delta PV$) varied with flow rate and injected volume, as shown in Fig. S1 of Supporting Information. Specifically, before water breakthrough during waterflooding, shorter $\Delta PV$ of 0.05 were used to capture more detailed dynamics due to the rapid breakthrough. In addition, individual image acquisition was performed at the precise breakthrough moments during both the oil drainage and waterflooding steps. For each experimental step, the total volume of injected fluid was kept constant (18 PV). However, a key issue to consider is that, for a given Q value, short stops in the flow were necessary to take the 12 local images before resuming fluid injection. The duration of this "stop-and-start" lapse time was, of course, adapted based on the corresponding Q. One issue with this approach is minimizing the impact of such short stopping times. To address this, separate experiments under the same conditions were performed by continuously injecting fluid, and the obtained data were compared to those from the "stop-and-start" procedure (see Fig. S2 in Supporting Information). Consequently, the stopping procedure was adjusted to ensure consistent results between both methods.

## 3 Results and Discussion

### *3.1 The Flow Dynamics in Oil Drainage and Waterflooding Steps*

In this paragraph we present the results of oil drainage and water flooding experiments prior to focus on tertiary recovery. The results of these experiments are contextualized with previously reported experiments for consistency.

In the oil drainage experiment (M = 80, Ca = 7.1×10⁻³), the full-chip image captured at oil breakthrough (**Fig. 4(a)**), shows a flat fluid front at macroscale, consistent with Saffman-Taylor theory [34], which predicts that interface perturbations are damped when the displacing fluid has a higher viscosity than the displaced fluid (Fig. S3 in Supporting Information). Otherwise, some localized fingers are observed at the front, as predicted by Lenormand et al. [35], since the flow regime is positioned within the transition zone (Fig. S4 in Supporting Information). This regime exhibits characteristics of both capillary fingering and stable displacement, resulting in breakthrough at 0.7 PV. It should be noted that all capillary numbers are here calculated using the convenient formula already proposed by Tang et al. [43], that considers the geometry and depth of 2D microfluidic chips:

$$Ca = \left(\frac{\mu v}{\sigma \cos\theta}\right) \underbrace{\frac{1}{k_r}\left(\frac{12}{2}\right)\left(\frac{D_t}{d_z}\right)^2\left(\frac{L_p}{D_t}\right)\frac{1}{\left(1-\frac{D_t}{D_b}\right)(\phi\varsigma)}}_{G} \qquad (1)$$

where, $\mu$ and $v$ are the viscosity and velocity of the displacing phase, respectively; $\sigma$ is the interfacial tension (IFT) between displacing phase and displaced phase; $k_r$ is the relative permeability, $d_z$ is the channel depth of the micromodel; $L_p$ is the length of one pore; $D_t$ is the characteristic pore‐throat width; $D_b$ is the characteristic pore‐body diameter; $\phi$ is the porosity; $\theta$ is the contact angle; and $\varsigma$ is an adjustable factor to relate permeability and pore geometry. The parameter values used in the present study are summarized in **Table S1** of **Supporting Information**, giving a geometric factor (G) of 450.

Anyway, details of the oil invasion process on pore scale were captured and illustrated in **Fig. 4(b)**, showing its dynamics as oil enters and advances through pore bodies and throats. When oil invades the pore body, it quickly saturates the larger spaces at $t_1$ (blue arrow) while slowly saturating smaller spaces. Then it remains blocked at the pore throat until the pressure within the pore body overcomes the critical capillary pressure, allowing it to advance at $t_2$ (red arrow), align with previous studies [3,29,31,49,73]. The remaining water is left in the form of a water film surrounding the pore surface (labelled ①), or as trapped water located in pore throats (labelled ②), corners (labelled ③), and in pores enclosed by small necks (labelled ④), as illustrated in **Fig. 4(c)**, consistent with reported results [3,73]. Water saturation ($S_w$) during the drainage showing excellent repeatability for the five repeated experiments (the blue error bar in **Fig. 4(d)**). Before oil breakthrough (BT~ 0.7 PV), $S_w$ rapidly decreased from 1 to 0.21 at a flow rate of 0.1 μL/min, forming a piston-like interface that swept the pores, leaving only a small amount of water behind the front, consistent with the findings of Lenormand et al. [35]. Afterward, $S_w$ decreased slightly, reaching 0.16 at steady state due to oil invading regions off the main flow path. Further sequential injection of 6 PV at flow rates of 0.5 μL/min and 1 μL/min reduced $S_{wi}$ to 0.13 and 0.11, respectively. These results indicate that flow rate has little effect on the remaining $S_w$ because most residual water remains in small spaces, as shown in **Fig. 4(c)**, which also confirmed by the water and oil distribution in the original pores in **Fig. 4(d)**, requiring higher pressure for effective removal.

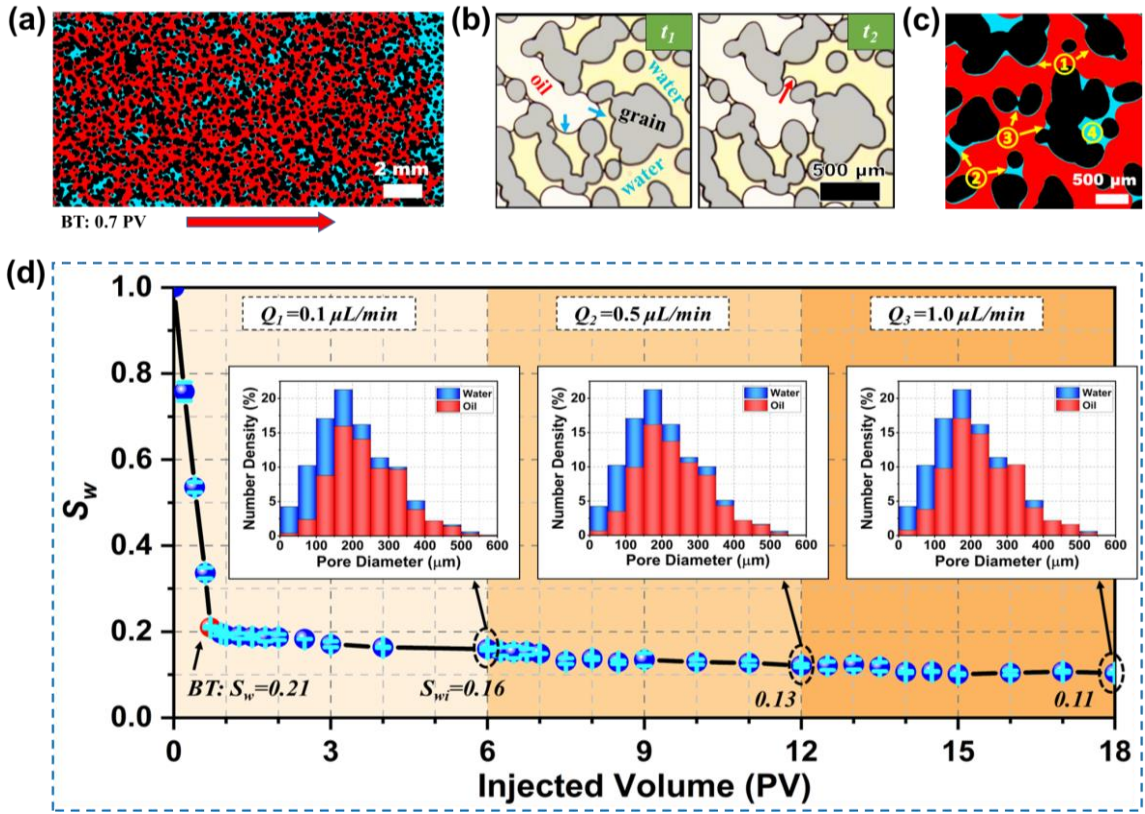

*Fig. 4. Results of drainage experiments. (a) Images of oil breakthrough during drainage, the bottom red arrow is the flow direction; (b) Snapshots of oil invading the pore bodies (blue arrows) and passthrough the pore throats (red arrows); (c) Forms of remaining water; (d) Variation of $S_w$ during the drainage experiments with three flow rates, and the corresponding water and oil distribution in original pores at the steady state for each flow rate.*

In the waterflooding experiment, breakthrough occurred at an injected volume of ~ 0.2 PV, with M=1/80 and Ca=8.9×10⁻⁵, as shown in **Fig. 5(a)**, exhibiting typical viscous fingering characteristics. In addition, a small amount of backflow was observed, indicating the presence of local capillary fingering (more details see Fig. S5 in Supporting Information), which aligns with the flow regime characteristics in the Ca-M phase diagram (see Fig. S4 in Supporting Information) [35]. The water saturation process, captured and displayed in **Fig. 5(b)**, shows water first invading small pores and quickly increasing saturation along the water films on the grain surface (blue arrows at $t_0$). As oil is displaced in the pores, pressure non-equilibrium at the pore throats causes water to spontaneously fill the throat, leading to the collapse and thinning of the oil (red ellipse, Δt = $t_1$-$t_0$ = 2 s) until snap-off occurs at $t_2$, (Δt = $t_2$-$t_1$ = 1 s), disconnecting the oil, which then moves along new pathways (red arrows), a process that described similarly by other literatures [28,39,74,75].

The oil saturation ($S_o$) was similar for the five repeated experiments. It decreased from an initial value of ~ 0.89 to a $S_{or}$ value of ~ 0.5 when water was injected at 0.1 μL/min, with BT occurring at ~0.2 PV and $S_o$ = 0.63. Subsequently, $S_{or}$ further decreased to 0.45 and 0.41 after sequential injection of 6 PV at flow rates of 0.5 μL/min and 1 μL/min, respectively, as shown in **Fig. 5(c)**. In a similar experiment, Hu et al. [38] have found that the invading saturation (that is the initial $S_o$ minus the actual $S_o$) between breakthrough and steady state during the viscous fingering flow regime exceeded 13%, consistent with our finding of ~13%. This higher recovery is attributed to the viscous fingering mechanism, where, after breakthrough, a substantial amount of oil remains trapped. With maintaining injection, the displacing fluid follows high-permeability pathways[32] and recovers oil along established routes. Full-chip images reveal that increasing the flow rate does not significantly induce residual oil breakup, as confirmed by the quantitative analysis of water and oil distribution within the original pores. This contrasts with findings of Haney et al. [76] in uniform microfluidic chips, where higher flow rates caused oil cluster fragmentation and size reduction. The difference likely arises from the irregular pore structure of our rock chip, which disperses the pressure increase and directs flow along existing low-pressure pathways, broadening flow paths without breaking oil clusters. In contrast, uniform pore structures allow pressure to distribute evenly, leading to random mobilization and fragmentation of oil clusters.

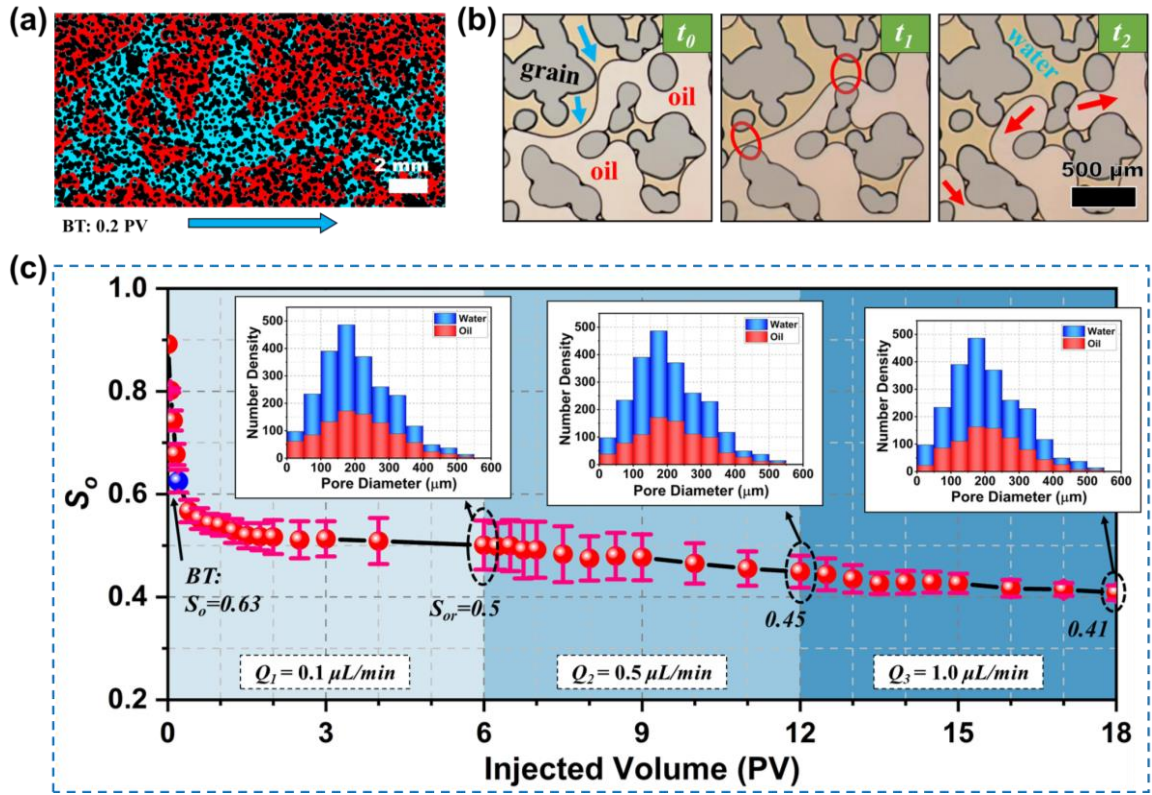

*Fig. 5. Results of waterflooding experiments. (a) Images of water breakthrough during waterflooding, the bottom blue arrow is the flow direction; (b) Snapshots of the water saturation (blue arrow), snap-off occurring at the pore*

*throat (red ellipse), and the afterward oil movement (red arrow); (c) Variation of oil saturation ($S_o$) during the waterflooding experiments with three flow rates, and the corresponding water and oil distribution in original pores at the steady state for each flow rate.*

### 3.2 Oil Clusters Mobilization During Mixture Flooding

The waterflooding experiment primarily explores the displacement mechanisms relevant to secondary recovery, providing limited insights into tertiary recovery. To address this gap, we conducted Glycerol/water mixture flooding experiments to investigate the influence of $\eta_R$ on recovery, residual oil cluster number, size and distribution during tertiary period. The dynamics and mechanisms of cluster mobilization were analysed both quantitatively and qualitatively. The viscosity of mixtures ranged from viscosity of water to that of oil, with the corresponding $\eta_R$ and Ca for each case shown in **Table 2**.

*Table 2 The value of $\eta_R$ and Ca for the five experimental cases*

|  | Case1 | Case2 | Case3 | Case4 | Case5 |
|---|---|---|---|---|---|
| **Glycerol concentration** (w/w) | 0 | 46 | 54 | 68 | 82 |
| $\eta_R$ | 1 | 4 | 8 | 20 | 80 |
| Ca | $8.89 \times 10^{-4}$ | $3.98 \times 10^{-3}$ | $6.95 \times 10^{-3}$ | $1.72 \times 10^{-2}$ | $7.09 \times 10^{-2}$ |

### 3.2.1 The Effect of Viscosity Ratio on Recovery Rate and Residual Oil Distribution

**Fig. 6(a)** shows global views of residual oil distribution in the microfluidic chip at steady state after 18 PV flooding with aqueous mixtures of different viscosity. The oil is represented in red, aqueous solutions in dark blue, and the solid grains in black. At $\eta_R = 1$ (continuously injecting water), a significant amount of residual oil remains, forming large and highly connected clusters. As the viscosity of the aqueous mixtures increases ($\eta_R$ increase), we clearly see the size and volume of residual oil to noticeably decrease. Moreover, the residual oil become more uniformly distributed across the microfluidic chip. At $\eta_R = 80$ (where the viscosity of mixture matches that of the oil), nearly all residual oil is mobilized, except for some oil trapped in dead-end pores. These results indicate that increasing the viscosity of the injected phase effectively reduces the size and volume of residual oil. Quantitatively, $S_o$ was evaluated during the injection of each mixture flooding case (see Fig. S6 in Supporting Information), and based on these $S_o$ results, the oil recovery in place was calculated as the ratio of the reduction in $S_o$ (initial $S_o$ minus the actual $S_o$) to the initial $S_o$, with the obtained results shown in **Fig. 6(b)**, where the initial recovery value is the waterflooding results of ~ 54%. As we can see, for $\eta_R = 1$, oil recovery changes only a little, as it resembles a typical waterflooding case. However, as $\eta_R$ increases, the steady state recovery improves significantly, with steady state being achieved earlier. When the total recovery is plotted against $\eta_R$, as shown in **Fig. 6(c)**, it is observed that the increase in recovery becomes less pronounced at high $\eta_R$, with recovery being more sensitive to $\eta_R$ in the range of $1 < \eta_R < 20$.

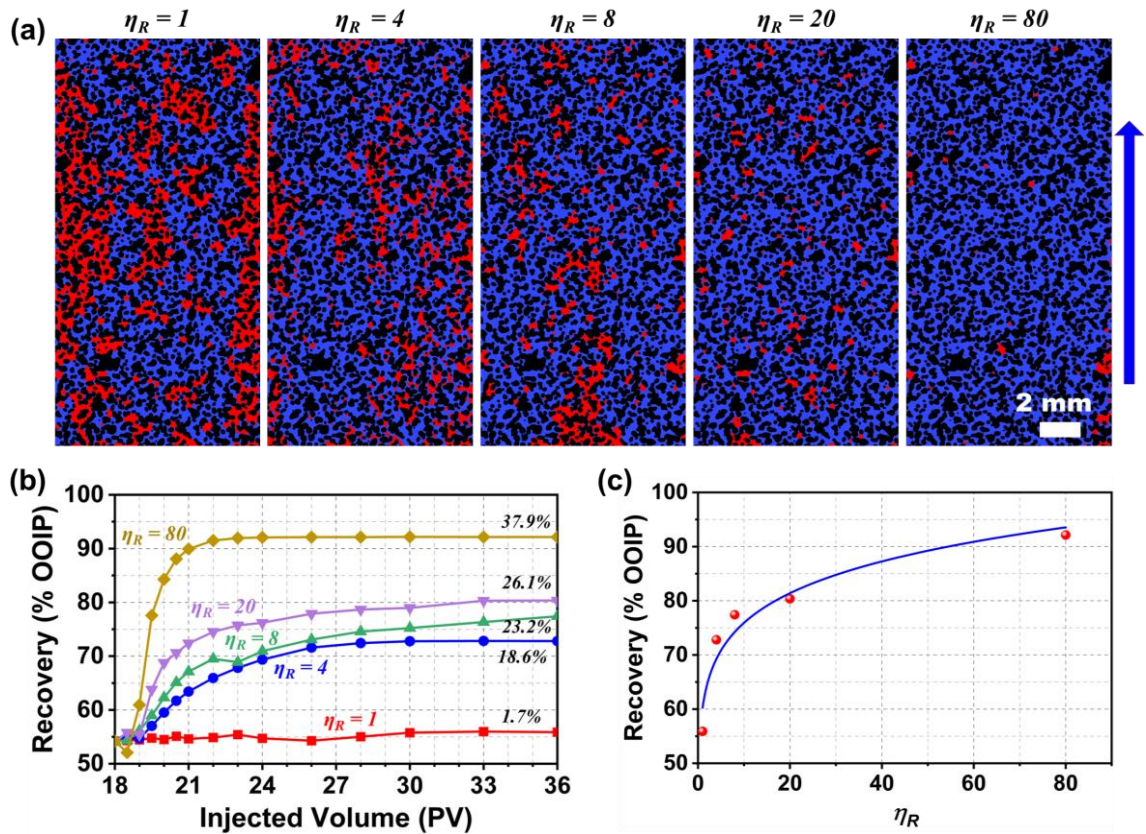

*Fig. 6. (a) Full-chip images at steady state after injecting 18 PV of Glycerol/water mixtures with five $\eta_R$; Blue arrow indicates flow direction. (b) Oil recovery as a function of injected pore volume with varying $\eta_R$; (c) Total oil recovery as a function of $\eta_R$.*

In other respects, we can quantitatively analyse the distribution of oil and aqueous solutions across the original pore sizes at steady state. As shown in **Fig. 7(a)** to **Fig. 7(e)**, the number of oil-filled pores decreases as $\eta_R$ increases. More importantly, as $\eta_R$ increases, the additional recovered oil initially comes from large and medium pores, then shifts to medium and small pores, highlighting how higher $\eta_R$ facilitates recovery across a broader range of pore sizes. This is further evidenced in **Fig. 7(f)**, where the continuous lines represent the residual oil pore size distribution for each $\eta_R$. It is clear that the peak of the distribution shifts towards smaller pore sizes as $\eta_R$ increases, with the oil in the larger pores predominantly being recovered.

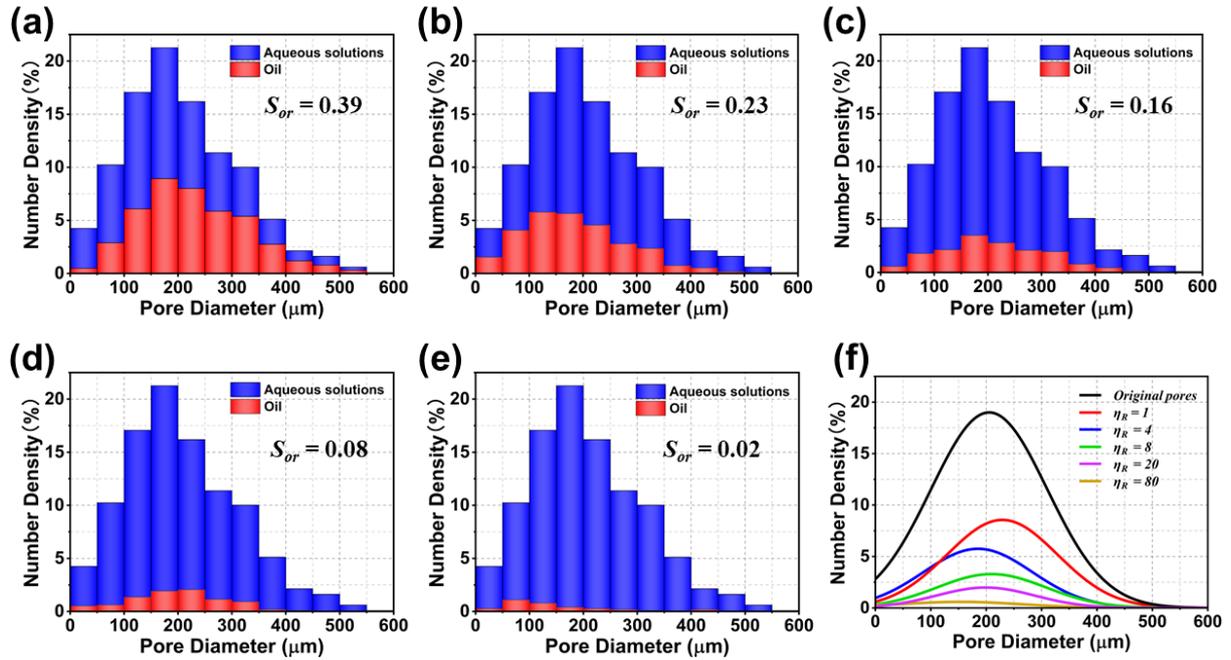

*Fig. 7. The distribution of oil and aqueous solution number density in the original pore size for five $\eta_R$: (a) $\eta_R=1$; (b) $\eta_R=4$; (c) $\eta_R=8$; (d) $\eta_R=20$; (e) $\eta_R=80$; and (f) distribution of total residual oil in the original pore size for five $\eta_R$ values.*

3.2.2 Dynamic of the Cluster Mobilization During Mixture Flooding

To more accurately and deeply understand the mobilization process of oil clusters, as shown in **Fig. 8**, clusters were categorized into droplets, blobs, and ganglia, following the literatures [24,77]. Droplets are smaller clusters typically located in the centre of pore bodies, either without solid contact or with one or two points of contact with the solid surface, blobs can fill an entire pore space and maintain some solid contacts, while ganglia are defined as larger oil clusters that may span more than one pores. Specifically, in data processing, droplets are defined as oil species with a size of 0-0.002 mm² and circularity between 0.7-1, blobs are characterized by a size range of 0.002-0.004 mm² and a circularity from 0.3-1, and ganglia are defined as oil species with a surface area larger than 0.004 mm² and a circularity from 0-0.7.

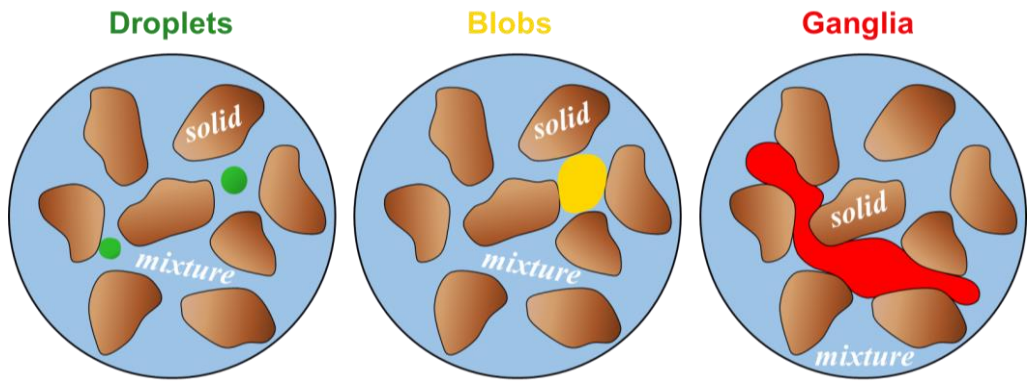

*Fig. 8. The schematic of three types of clusters, represented in different colours: green for droplets, yellow for blobs, and red for ganglia.*

As an illustrative example, **Fig. 9(a)** and **Fig. 9(b)** shows the number and total volume in the populations of droplets, blobs, and ganglia during mixture flooding for the $\eta_R$=4 case (Full-chip images are displayed in Fig. S7 of Supporting Information). Droplets were the most numerous, followed by blobs, with ganglia being the least frequent. In contrast, the total volume distribution was reversed: ganglia occupied the largest volume, followed by blobs, with droplets contributing the least. Consequently, this demonstrates that ganglia, as the largest contributors to the total volume, play a primary role in determining $S_o$. Their volume reduction directly decreases $S_o$, while blobs and droplets, despite their higher numbers, have a relatively limited impact.

In details, during the initial period of mixture flooding, the injected mixture firstly displaces water and initiates a lateral sweep of the chip, leading to the rapid breakage of large ganglia. This breakage forms smaller clusters, including droplets, blobs, and smaller ganglia, resulting in an increase in their numbers (before 2 PV), as shown in **Fig. 9(a)**. Simultaneously, the total volume of ganglia decreases significantly, while the volumes of droplets and blobs slightly increase, as shown in **Fig. 9(b)**, highlighting the impact of mixture flooding on cluster dynamics during this period. Once breakage occurs, droplets and blobs become candidates for removal, as they are barely subjected to capillary forces [27], they are gradually mobilized and are then transported out of the system, reducing their total number over time. In other words, the characteristic time for oil transport is longer than that required for ganglia breakage under our experimental conditions. Therefore, as $S_o$ decreases to a critical value, the rate of ganglia breakup eventually falls below the rate at which small clusters are displaced from the microfluidic chip, causing the removal mechanism to slow down. This marks a transition around 2 PV, leading to a gradual decline in the number and volume of all clusters. This trend is similar to the findings of Li et al. [44] during secondary displacement, where snap-off generates mobile dispersed droplets, resulting in an initial increase followed by a decrease in the number of non-wetting phase clusters. However, our study delves further into tertiary recovery by analysing the mobilization dynamics of these three types of residual oil clusters (droplets, blobs, and ganglia), quantitatively tracking changes in their number, occupied volume, and location over higher injection volumes.

By dividing the total volume by the number of each cluster type, the average volume per cluster is obtained, as shown in **Fig. 9(c)**. Therefore, we can see that the average volumes of blobs and droplets are almost constant, likely due to their small size and regular shape making them less affected by shear stress. In contrast, ganglia average volume briefly increase in the vicinity of 0.5 PV, because at short times when ganglia breakage begins, smaller clusters are produced, and concomitantly, blobs and droplets may merge with these clusters rather than being immediately removed [27] during this period. Afterward, ganglia continue breaking up, significantly reducing their average volume and feeding the medium by new droplets and blobs that are transported away by the main flow. After 2 PV, the average volume of ganglia change only smoothly as the mixture fully sweeps the microfluidic chip. In this view, such a process occurs because the mixture first has to displaces water along existing pathways before it sweeps other regions, and ganglia breakage rate is maximum at approximately 2 PV.

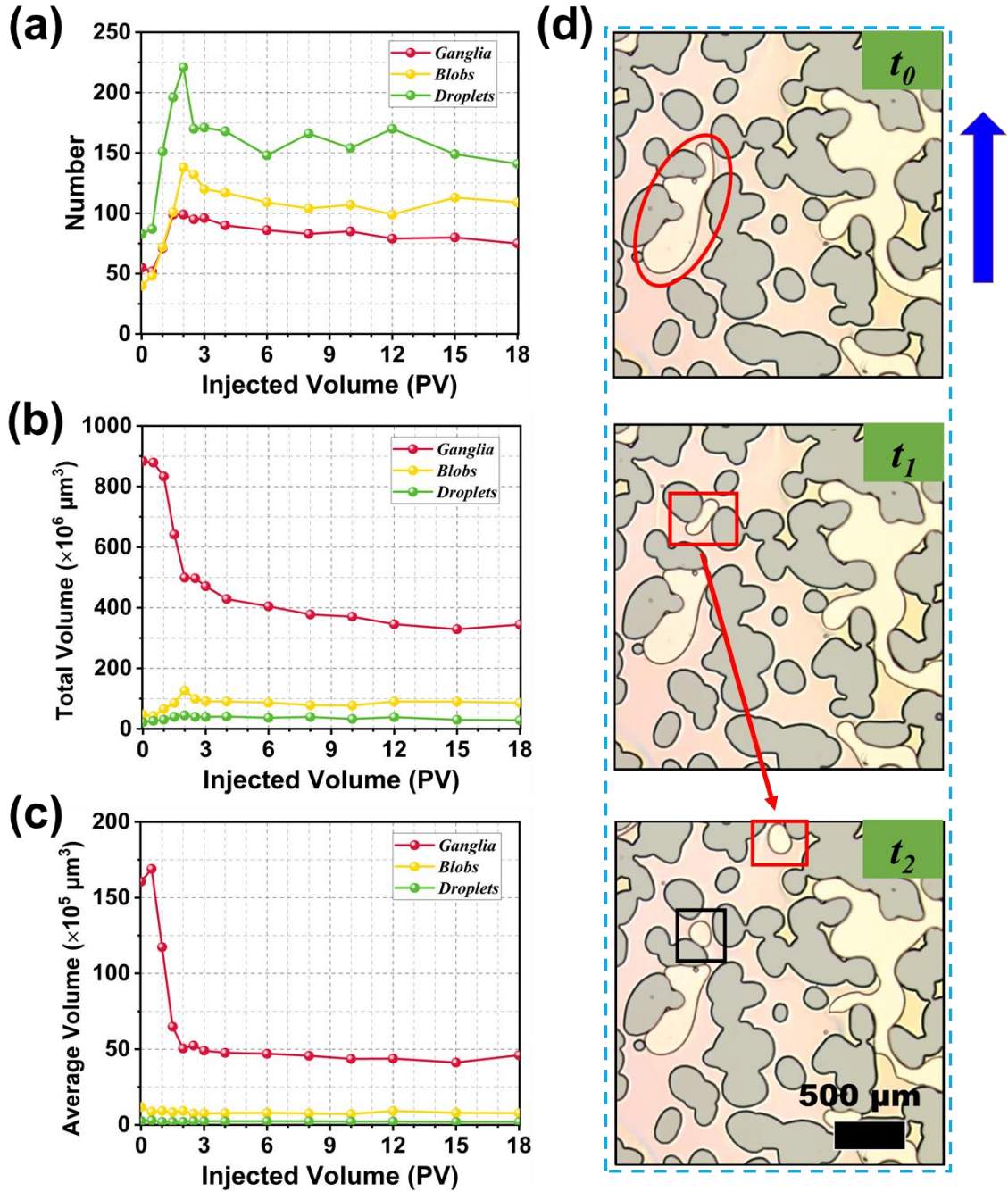

*Fig. 9 Changes of three clusters during mixture flooding for $\eta_R = 4$ with: (a) Number; (b) Volume; and (c) Average volume. (d) Snapshots of ganglia mobilization dynamic during mixture flooding, where light yellow is oil, light pink denotes the mixture, and grey corresponds to solid grains. Blue arrow indicates flow direction.*

A more direct evidence come from local snapshots that were taken at successive time intervals (see **Fig. 9(d)** and more details shown in Supplementary Video). In this figure, a large ganglion (red ellipse) undergoes rapid breakup at the pore throat within 1 s at $t_1$, forming small blobs and droplets (red box) that move through the pore space, reaching the edge of the field of view 2 s later at $t_2$. The left ganglion remains trapped but continues breaking up under sustained flooding, generating new blobs and droplets (black box) until it is fully mobilized. Unlike studies on oil cluster mobilization driven by increasing the Ca in secondary recovery methods, our findings reveal a distinct mechanism in tertiary recovery when fluids of monitored viscosity are injected after the waterflooding step. While some cluster merging was observed, the dominant process is the fragmentation of large ganglia and born blobs and droplets are then removed, making the generation and transport of small oil clusters the primary mechanism for residual oil mobilization, as it was hypothesized earlier by Haney et al [76]. This come in contrast with the classic scenario that describe the recovery process as a whole mobilization of all oil clusters whatever their actual size and structure [27,78].

Let us focus now on an isolated ganglion (see **Fig. 10**) and calculate the capillary pressure $P_c$ inside it before and after its breakage, following the method of Alzahid et al. [74]. Before its breakage, the capillary pressure within the ganglion is seen to decrease from $P_{c2}$ upstream to $P_{c1}$ downstream (data detailed in **Table 3**). When the Glycerol/water mixture is then injected, both shear forces and the pressure gradient increases, causing the ganglion to thin and elongate at the pore throat (yellow arrows in **Fig. 10(a)**). When the pressure exceeds the critical capillary force there, the ganglion breaks into a small part "B" downstream (see **Fig. 10(b)**) that may be easily mobilized, and a larger part "A" upstream that remains temporarily trapped under dominant capillary forces unless further breakup occurs. At breakup, the throat pressure exceeds $P_{c1}$ and $P_{c2}$ (see **Table 3**), triggering rapid withdrawal movement (white arrows in **Fig. 10(b)**) and redistributing stresses, leading to increased interfacial curvature in the new segments. An analogous mechanism was evidenced by Zarikos et al. [24], who investigated the breakage mechanism of ganglia under flow in a 2D packed glass bead medium. They measured the velocity field inside trapped ganglia using the microscopic particle tracking velocimetry technique and found that, under wetting-phase pressure, internal vortices formed within the ganglia, which flattened and thinned at locations of throat where the velocity was almost zero, before potentially breaking.

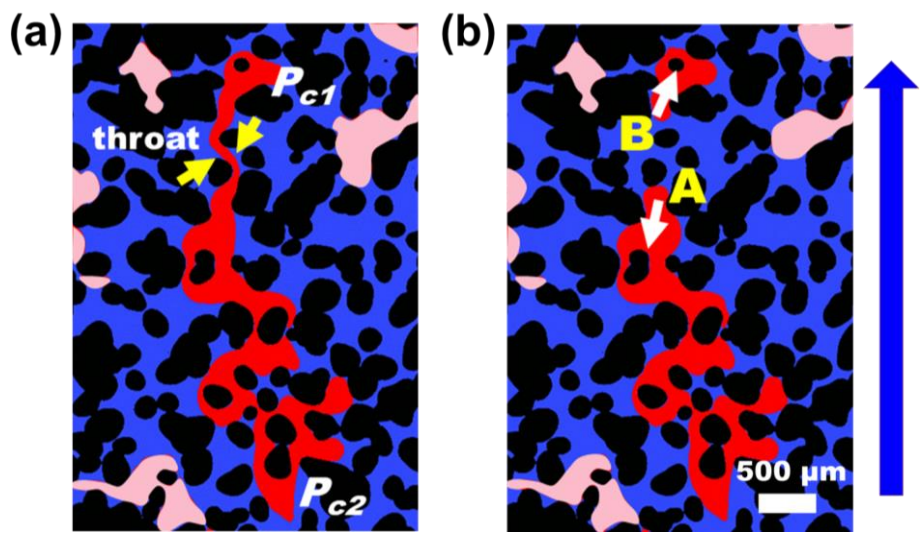

*Fig. 10. Isolated ganglion breakage at the pore throat: (a) Before breakup, and (b) After breakup. Solid grains are shown in black, the analysed oil ganglia in red, other remaining oil in light red, and the Glycerol/water mixture in dark blue. Blue arrow is flow direction.*

**Table 3** *Capillary pressure at throat and $P_{c1}$ and $P_{c2}$ at the moment of breakup*

| Location | IFT (N/m) | Curvature diameter (m) | $P_c$ (Pa) |
|---|---|---|---|
| **Throat** | $5.5\times10^{-2}$ | $1.15\times10^{-4}$ | $9.57\times10^{2}$ |
| $P_{c1}$ | $5.5\times10^{-2}$ | $2.34\times10^{-4}$ | $4.7\times10^{2}$ |
| $P_{c2}$ | $5.5\times10^{-2}$ | $1.86\times10^{-4}$ | $5.91\times10^{2}$ |

*3.2.3 Influence of mixture viscosity on oil cluster removal*

To summarize, we investigated in detail the breakage and removal of oil clusters remaining after waterflooding when a more viscous Newtonian fluid was injected. As described in the procedure section, these experiments were conducted using Glycerol/water mixtures of varied viscosities to evaluate how $\eta_R$ impacts the recovery of additional oil left in place after waterflooding. After injecting 18 PV of the mixture, microfluidic chip images were processed to classify clusters into ganglia, blobs, and droplets, represented in different colours for clarity. **Fig. 11(a)** shows synthetic images of these clusters for all $\eta_R$ values at steady state, with solids excluded and the pore structure disregarded.

Qualitatively, for $\eta_R = 1$ (continuous waterflooding), the remaining oil primarily consists of large ganglia. As $\eta_R$ increases, these ganglia gradually disappear, reducing their overall size with a more uniform size distribution. From a quantitative point of view and as it is shown in **Fig. 11(b)** and **Fig. 11(c)**, most of the additional recovered oil originates from ganglia breakage, which total volume decreases sharply with $\eta_R$. Hence and although ganglia are few, given their large size any breaking event has a significant reduction of $S_{or}$ and have therefore the most determinant effect of oil sweeping enhancement.

To go further in a detailed description of the oil recovery enhancement by Glycerol/water mixture injection, the oil clusters removal mechanism may be phenomenologically described as follows:

Before mixture injection, oil clusters remain integer and immobile throughout the chip. Once the mixture is introduced at the same flow rate, it first replaces water upstream, then flows preferentially along pre-existing waterflooding pathways, displacing water and gradually increasing pressure within the chip. This pressure build-up rapidly stresses the clusters, breaking them into smaller droplets and blobs, which are subsequently transported. The elevated pressure also mobilizes and redistributes downstream clusters, altering their size and location. As injection progresses and more areas of the chip are swept, further ganglia breakup occurs, with the fragments carried away by the mixture. In this picture, the characteristic time for ganglia breakup is shorter than that for small oil fragments transport, enabling efficient mobilization. Overall, oil cluster evolution before mixture breakthrough is dynamic and is related to local change of fluid composition and flow structure in a cascading pressure build-up, local increase of stress, oil clusters breakage and mobilization and transport.

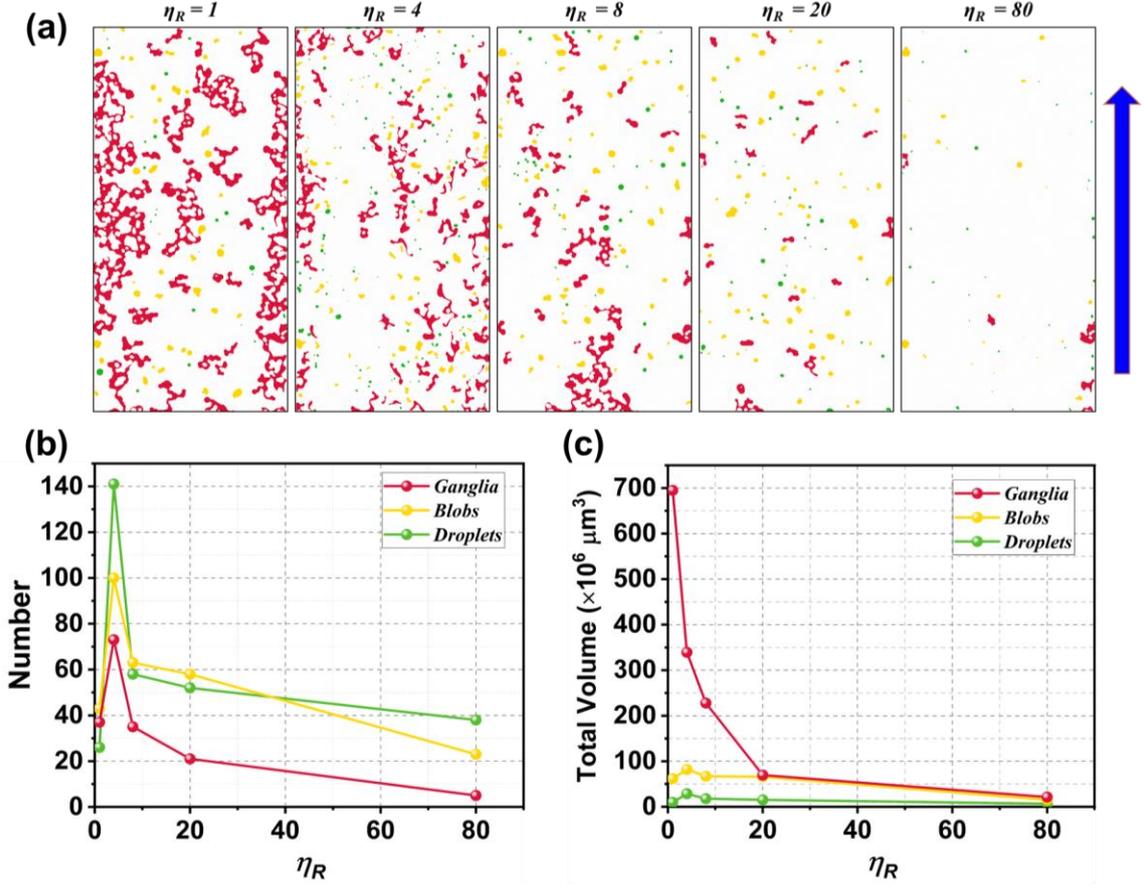

*Fig. 11. (a) Full-chip image showing the location of ganglia (red), blobs (yellow), and droplets (green) at steady state for five $\eta_R$. Blue arrow is flow direction. (b) Corresponding numbers and (c) volumes of ganglia, blobs, and droplets as a function of $\eta_R$.*

*3.3 Cluster Size Distribution and Capillary Desaturation Curve*

Through the analysis here above, we discussed the mobilization dynamics of oil clusters depending on whether they are droplets, blobs or ganglia. However, when a statistic is formed by encompassing all of these kind of oil clusters, their normalized number *N(s)* should depend on their normalized size *s* following a power law as:

$$N(s) \sim s^{-\tau} \qquad (2)$$

where $\tau$ is a critical scaling exponent, it holds significant physical implications: a smaller $\tau$ indicates larger residual clusters, which are easier to mobilize. Here, the normalized *N(s)* is defined as the number of ganglia with a given surface area (in pixels) divided by the number of the smallest ganglia, while normalized size *s* is the surface area of a ganglion divided by that of the smallest one [79]. Several authors have reported $\tau$ values ranging from 1.8 to 2.05 [78–81]. Such a power law is obviously convenient to fit the data plotted on **Fig. 12(a)** where the relationship $N(s) \cong s^{-2}$ is drawn to serve as a guide for eyes noting by the way that the data cloud is large at high s values as every cluster there are fewer. However, Iglauer and Wülling [81] have shown and discussed how the extent of size interval does influence the value of the exponent $\tau$ that may in turn be much less than 2 and even approaching unity for very large size intervals.

To complete the current analysis, the capillary desaturation curve (CDC) was constructed by representing $S_{or}$ versus the capillary number. The data used for this analysis included those obtained from experiments where the viscosity of the injected aqueous fluids was varied, as well as those from flow rate changes during the waterflooding period (see **Fig. 12(b)**). The resulting CDC is therefore seen to merges into a single trend as both the viscosity change and the flow rate variation concurrently impact the value of capillary number. In particular, $S_{or}$ decreases as Ca increases beyond the critical threshold of Ca ≈ $10^{-4}$, above which viscous stress becomes high enough to induce oil clusters breakage and transport, enhancing then the oil sweeping efficiency, in accordance with the finding of other authors who have used both Newtonian fluids and shear thinning polymeric fluids for which the viscosity was dependent on the effective shear rate actually experience [1,42,82].

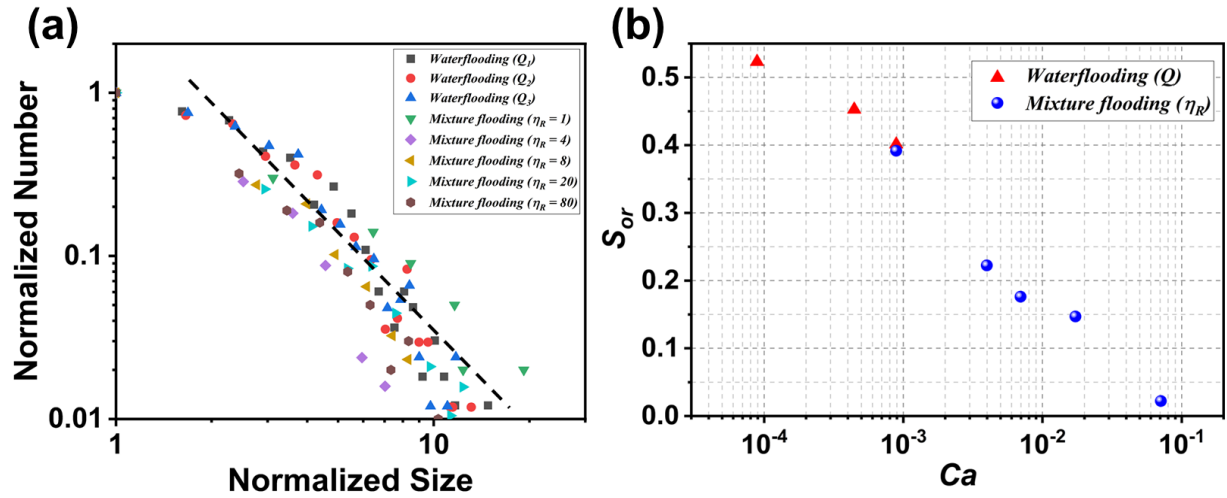

*Fig. 12. (a) Normalized number of clusters versus their normalized size for different Q and $\eta_R$ values; the dashed line represents the power law $N(s) \cong s^{-2}$. (b) The capillary desaturation curve showing $S_{or}$ variation in function of Ca.*

## 4. Conclusions

The objective of the present work was to investigate the influence of viscosity ratio on oil mobilization in tertiary EOR by means of microfluidic setup that allow pore scale direct visualization and characterization of diphasic flow in water-wet microfluidic chips. To summarize, the following key behaviours may be highlighted:

(1) Under our experimental conditions, as oil was injected into a water saturated chip, the invasion was approaching a piston-like displacement with a BT of 0.7 PV. The remaining water saturation at steady state was also seen to decrease only a little when the flow rate is increased.

(2) If water was subsequently injected, an early BT was observed at ~0.2 PV, due to the viscosity contrast that induces a viscous fingering flow structure with large oil clusters left behind in particular locations. Description of mechanisms involved in both drainage and waterflooding was presented at macroscopic scale as well as at pore scale with the help of acquired images, in the course of these processes showing characteristic flow structures and how oil clusters breakage and transport take place.

(3) In tertiary oil recovery step, Glycerol/water mixtures effectively enhance porous medium sweeping, with higher viscosity ratio leading to greater oil recovery with the remaining oil in place being smaller with a narrower distribution. Oil clusters are categorized in droplets, blobs and ganglia based on their size and curvature. Ganglia, despite being the least numerous, account for a large proportion of the total volume, whereas blobs and droplets contribute only a small proportion. When experimental results are presented in terms of global cluster size distribution or in terms of the capillary desaturation curve as usual, they are shown to be in accordance with expectations. More interesting are the outcomes when the mobilization mechanisms are analysed in details. Indeed, we observed that during mixture flooding, ganglia break into smaller ganglia, blobs, and droplets that are subsequently mobilized and transported away, while remaining parts of the original ganglia remain trapped. Thus, while the number of ganglia changes slightly with increasing viscosity of the injected mixture due to breakage, their total volume decreases significantly, thereby lowering the oil saturation. In contrast, the total volumes of blobs and droplets is seen to only weakly change by the increase of mixture viscosity and even their number may temporarily increase as they result from ganglia breakage. Therefore, we evidenced that the process could be separated in two main steps: ganglia (and probably few big oil blobs) breakage that feed the medium in blobs and droplets, and second, transport of these mobile oil clusters, with a characteristic time which is longer than that required for ganglia breakage.

Naturally, these experiments require further improvements. As perspective of the presented study we plan to enhance the representatively of the experimental protocol by using larger intervals of viscosity ratio and flow rates and chips of non-uniform depth. We also plan to use the developed experimental pipeline to quantitatively characterize the sweeping potential of viscoelastic fluids.

## Author contributions

A. Omari and G. Sciumè offered the idea and designed the experiments. H. Pi performed the experiments and analysed experimental data. A. Omari and G. Sciumè helped to data analysis and provided guidance. The original draft of the manuscript was written by H. Pi, with subsequent review and editing by A. Omari and G. Sciumè.

## Declaration of Competing Interest

The authors declare that they have no known competing financial interests or personal relationships that could have appeared to influence the work reported in this paper.

## Data availability

No data was used for the research described in the article.

## Acknowledgments

This work was supported by the financial assistance of the China Scholarship Council (No. 202008110244, Recipient: Haohong PI) during PhD research. It also acknowledges the additional support provided by the *Institut Universitaire de France* (IUF) through Giuseppe Sciumè.

## Statement

During the preparation of this work the authors used ChatGPT from time to time in order to improve the readability and language of the manuscript. After using this tool/service, the authors reviewed and edited the content as needed and take full responsibility for the content of the published article.